\documentclass[onecollarge]{svjour2}       % onecolumn "king-size"
\smartqed  % flush right qed marks, e.g. at end of proof
\usepackage{graphicx}
\usepackage{natbib}

 \journalname{3D RESEARCH, Digital Holographic Microscopy issue}
\begin{document}

\title{3D exploration of light scattering from live cells in the presence of gold nanomarkers using holographic microscopy %\thanks{Grants or other notes
}


\author{Fadwa Joud         \and
        Nilanthi Warnasooriya \and
        Philippe Bun\and
        Fr\'{e}d\'{e}ric Verpillat \and
        Sarah Y. Suck\and
        Gilles Tessier\and
        Michael Atlan\and
        Pierre Desbiolles\and
        Ma\"{\i}t\'{e} Coppey-Moisan\and
        Marie Abboud\and
        Michel Gross\and
        %etc.
}


\institute{F. Joud \at
              Laboratoire Kastler Brossel, \'{E}cole Normale Sup\'{e}rieure, 24 Rue Lhomond, 75005 Paris, France \\
              Tel.: +33-1-44-32-35-12\\
              Fax: +33-1-44-32-34-34\\
              \email{fadwa.joud@lkb.ens.fr}           %  \\
           \and
           N. Warnasooriya \at
              Institut Langevin, ESPCI, 10 Rue Vauquelin, 75005 Paris, France\\
              \emph{Present address: Dept. of Biomedical Engineering, Texas A\&M University, College Station, TX 77843, U.S.A.}
           \and
           P. Bun \at
              Institut Jacques Monod, Universit\'{e} Paris-Diderot, 15 Rue H\'{e}l\`{e}ne Brion, 75205 Paris, France
           \and
           F. Verpillat \at
              Laboratoire Kastler Brossel, \'{E}cole Normale Sup\'{e}rieure, 24 Rue Lhomond, 75005 Paris, France
           \and
           S. Y. Suck \at
              Institut Langevin, ESPCI, 10 Rue Vauquelin, 75005 Paris, France\\
              Fondation Pierre-Gilles de Gennes pour la recherche, 29 Rue d'Ulm, 75005 Paris, France
           \and
           G. Tessier \at
              Institut Langevin, ESPCI, 10 Rue Vauquelin, 75005 Paris, France
           \and
           M. Atlan \at
              Institut Langevin, ESPCI, 10 Rue Vauquelin, 75005 Paris, France
           \and
           P. Desbiolles \at
              Laboratoire Kastler Brossel, \'{E}cole Normale Sup\'{e}rieure, 24 Rue Lhomond, 75005 Paris, France
           \and
           M. Coppey-Moisan \at
              Institut Jacques Monod, Universit\'{e} Paris-Diderot, 15 Rue H\'{e}l\`{e}ne Brion, 75205 Paris, France
           \and
           M. Abboud \at
              D\'{e}partement de Physique, Facult\'{e} des Sciences, Universit\'{e} Saint-Joseph, Beirut, Lebanon
           \and
           M. Gross \at
              Laboratoire Kastler Brossel, \'{E}cole Normale Sup\'{e}rieure, 24 Rue Lhomond, 75005 Paris, France and Laboratoire Charles Coulomb  - UMR 5221 CNRS-UM2 Universit\'{e} Montpellier II place Eug\`{e}ne Bataillon 34095 Montpellier
}

\date{Received:  / Accepted: }

\maketitle

\begin{abstract}

In this paper, we explore the 3D structure of light scattering from
dark-field illuminated live 3T3 cells
in the presence of 40 nm gold nanomarkers. For this purpose, we use a
high resolution holographic microscope combining the off-axis
heterodyne geometry and the phase-shifting acquisition of the digital
holograms. A comparative study of the 3D reconstructions of the
scattered fields allows us to locate the gold markers which
yield, contrarily to the cell structures, well defined bright
scattering patterns that are not angularly tilted and clearly located
along the optical axis (z). This characterization is an unambiguous
signature of the presence of gold biological nanomarkers, and validates
the capability of digital holographic microscopy to discriminate them
from background signals in live cells.

\end{abstract}

\section{Introduction}
\label{intro}

Gold nanoparticles attract great scientific and technological interest because of their
physical and chemical characteristics. In particular, the optical tracking of gold
nanoparticles in biology have gained popularity for several reasons. Gold nanoparticles
provide high scattering efficiencies (\cite{jain2006calculated}) and they can be detected
directly using dark field or total internal reflection (TIR) illumination down to
particle diameters of 40 nm as shown by \cite{sonnichsen2000spectroscopy}. Unlike
fluorescent markers, they are immune to photo bleaching, and they are potentially
non-cytotoxic (see \cite{west06}). Because of these properties, the use of gold nanoparticles as
biomarkers for live cell imaging using photothermal tracking (see \cite{cognet2002},
\cite{boyer2003} and \cite{lasne2006}) has a high potential.

As shown by \cite{atlan2008heterodyne}, holography has  proved its ability to image and
localize  gold nanoparticles in 3D, either for fixed particles spin coated on a glass
substrate or in free motion within a water suspension. More recently,
\cite{absil2010photothermal} have shown that heterodyne holography also allows the
photothermal detection of 10 nm gold particles, and \cite{warnasooriya2010imaging} have
imaged 40 nm gold particles in a cellular environment. In that last experiment the
particle holographic signal is superimposed with the light scattered by the cell refractive
index fluctuations, which yield a speckle field. For particles imaging, this induced speckle
is a parasitic signal, but in many other situations, like in Dark Field microscopy , or
in Differential Interference Contrast (DIC) microscopy (see \cite{goldberg1986stages} )
this speckle is the main source of contrast that is used to image the cell itself. It is thus important
to discriminate the particle signal from the cell parasitic speckle.

In this paper, we  have imaged biological samples (3T3 cells) labeled with 40 nm gold
particles using the digital holographic setup described in  \cite{warnasooriya2010imaging}. We
have performed the 3D holographic reconstruction of the wave-field scattered by the
samples, which are illuminated at 45$^\circ$ in a total internal reflection
configuration, and we have shown that these wave-fields  are noticeably different for
the particle, and the speckle signal. We showed here that important information can be derived not only from the intensity of the bright spots caused by the gold particles, but also from the 3D shape of the light scattering pattern, which is easily accessed using holography. We showed, in particular, that the speckle
signal keeps memory of the illumination direction, while the particle signal does not. This
result has been confirmed by imaging samples of cells that have not been labeled with gold particles, and  samples of free gold particles. The shape of the wave-field scattered by the sample can thus be
used as a signature that helps to discriminate the particle signal from the speckle.

\section{Materials and Methods}
\label{sec:1}

\subsection{Biological specimen preparation}
\label{sec:2}

\begin{figure}
%\begin{center}
   \begin{tabular}{c}
   \includegraphics[height=6cm]{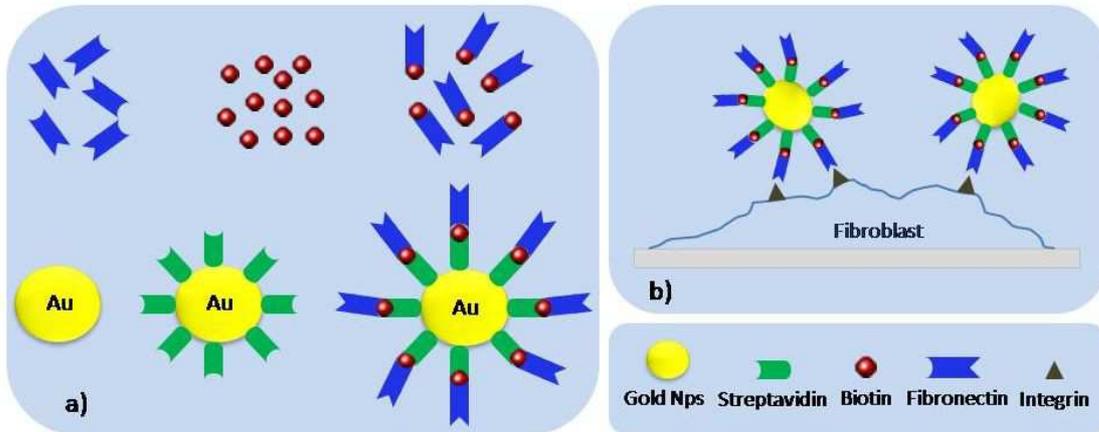}
   \end{tabular}
  %\end{center}
\caption{Gold bioconjugates and fibroblasts preparation procedures. (a) Gold bioconjugates functionalisation. (b) Fibroblasts-gold nanoparticles coupling.}
\label{fig:1}       % Give a unique label
\end{figure}

The biological specimens that we imaged are monolayers of live NIH 3T3 mouse fibroblasts labeled
with 40 nm gold particles via their integrin cellular surface receptors. Streptavidin-coated gold
nanoparticles were attached to the surface cellular integrin receptors via biotin and
fibronectin proteins: see Fig.\ref{fig:1}. Streptavidin and biotin are very well known
for their strong affinity towards each other, and fibronectin, an extracellular matrix
protein, has the property of interacting specifically with cellular surface receptors of
integrin family.

Fibronectin  proteins  (fibronectin from bovine plasma, Sigma, St Louis, MO) were biotinylated using EZ-Link$^{\textcircled {\scriptsize R}}$Sulfo-NHS-LC-Biotin
according to the provider protocol (Pierce, Rockford, IL). The final concentration of
biotinylated-fibronectin solution was 0.447 mg/mL. The
streptavidin-coated gold conjugates of 40 nm average diameter (BioAssay, Gentaur, France) were rinsed twice with 1X PBS (Phosphate Buffered Saline) (pH~=~7.25). We than diluted 10 $\mu$L of the gold solution in
990 $\mu$L of the same PBS buffer solution. Then the dilute gold solution
was incubated with 50 $\mu$L of the biotinylated-fibronectin solution for four hours at room
temperature to allow the specific streptavidin-biotin bonding. The final functionalized gold particles solution was
kept at 4$^\circ$C and used within 24 hours after its preparation in order to ensure
maximum functionality. Before every use, the functionalized gold particles solution was sonicated.

48 hours before the observation, monolayers of 3T3 cells were cultured   in Duelbecco's modified Eagle's medium (DMEM Gibco, Invitrogen, Carlsbad, CA) supplemented with 10\% fetal calf serum (PAA Laboratories GmbH)  on 32 mm
diameter fibronectin-coated glass cover slips (fibronectin from bovine
plasma, Sigma, St Louis, MO) at 37$^\circ$C in a 5\% CO2 atmosphere. After 24 hours of incubation, we added to each coverslip a solution composed of DMEM (2 mL) plus 500 $\mu$L of the functionalized gold particles solution. The binding
of integrin and fibronectin occurs at this level allowing the cells to attach, on their
surface, the functionalised gold nanoparticles.

The coverslip  containing adherent 3T3 cells tagged with 40 nm gold nanoparticles was
mounted on a specific observation chamber. In order to maintain the physiological pH condition during the experiments, cells were kept in DMEM-F12 medium (DMEM-F12 without Phenol red, B12 vitamin, Riboflavin, 0.5\% fetal calf serum
and supplemented with 20 mM of HEPES [(4-(2-hydroxyethyl)-1-piperazineethanesulfonicacid)] and L-Glutamine from PAA Laboratories). We measured the level of biotin incorporation  on an HABA [2-(4' -Hydroxyazobenzene)  Benzoic Acid)] quantitation assay to verify the efficiency of the biotinylation protocol. Average number of biotin molecules obtained per fibronectin is 2.5.

\subsection{Holographic Microscope Experimental Setup}
\label{sec:3}

\begin{figure}
%\begin{center}
   \begin{tabular}{c}
   \includegraphics[height=7.2cm]{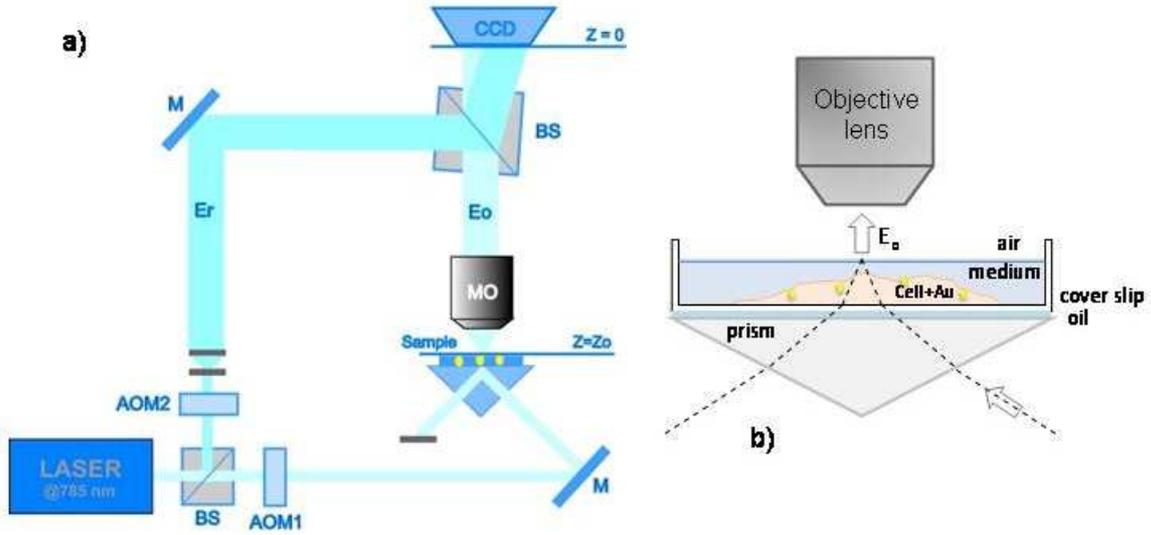}
   \end{tabular}
  %\end{center}
\caption{(a) Experimental setup : AOM1, AOM2: acousto-optic modulators; M: mirror; MO:
microscope objective (NA = 0.5);  BS: beam splitter; PBS: polarizing
beam splitter; CCD: CCD camera; $E_r$: reference field; $E_o$:
scattered field;  z = 0: CCD plane; z = $z_0$:
CCD conjugate plane with respect to MO.
(b) Details of the total internal reflection optical arrangement
that is used for dark-field illumination of the biological sample.}
\label{fig_2setup}       % Give a unique label
\end{figure}

Fig.\ref{fig_2setup} illustrates the optical setup. The illumination source is a
single-mode near infrared laser diode emitting at $\lambda =785$nm (DL7140-201S 80 mW
Laser Diode $@$90 mA current). A polarizing beam splitter cube (PBS) is used to split the
original illumination laser light into two beams, a reference beam (complex field $E_R$,
frequency $f_R$) and an object illumination beam (complex field $E_0$, frequency $f_0$)
forming the two arms of a Mach-Zehnder interferometer. A combination of a half wave
plate  and two neutral density filters is used to prevent the saturation of the detector
by controlling the optical power traveling in each arm. Two acousto-optic modulators
(AOM1, AOM2) driven around 80 MHz and using the first order of diffraction,
shift both frequencies at respectively $f_{AOM1}$ and $f_{AOM2}$ .

The object beam illuminates the sample, prepared as described in Section \ref{sec:2}, by
provoking total internal reflection (TIR) at the medium-air interface in order to prevent direct
illumination light from entering the system. The evanescent
wave locally frustrated and the illumination wave directly scattered by the beads and cells give off a propagating scattered wave
(complex field $E$), which is collected by a microscopic objective (MO, $50 \times$
magnification, NA=0.5, air). A beam splitter is then used to combine the scattered object
wave and the reference wave which is slightly angularly tilted ($ \sim 1^\circ$) with
respect to the propagation axis of the object wave in an off-axis configuration. A half
wave plate  on the object illumination arm aligns the polarization of the corresponding
beam ensuring its optimal interference with the reference beam. A CCD camera (Roper Cascade 512F,
$512 \times 512$ square pixels of 16 $\mu$m size, exposure time 100 ms, frame rate
$f_{CCD}$ = 8 Hz) detects the interference pattern (hologram) and sends it to a computer.
The hologram is then numerically treated and the complex field $E(x,y,z)$ is
reconstructed numerically.

\subsection{Holographic acquisition}
\label{sec:acquistion}

In order to filter out unwanted parasitic signals, we use a heterodyne modulation. A four-phase demodulation method is used  to record holograms. This method consists in
acquiring a sequence of images with a relative phase shift $\Delta\varphi=\pi/2$ between
two consecutive frames. The reference wave is frequency shifted  by tuning the two
acousto-optic modulators AOM1 and AOM2 as done by \cite{Leclerc2000}, and we get, as shown by \cite{atlan2007aps}, an accurate phase shift $\Delta\varphi$ that
simplify the phase shifting digital holography demodulation. The heterodyne beat
frequency is thus:
\begin{equation}\label{Eq_1a}
    \Delta f={f}_{AOM1}-{f}_{AOM2}=\frac{f_{CCD}}{4}
\end{equation}
where $f_{CCD}=$8Hz is the frame rate frequency of the CCD camera. The camera records a
sequence of 32 frames $\emph{I}_{0}$,...,$\emph{I}_{31}$, and the object field $\emph{E}$
on the CCD plane ($z=0$) is given by:
\begin{equation}\label{eq2}
    \emph{E}(x,y,z=0) = \sum_{n=1}^{M}  j^{~n} \emph{I}_{n}
 \end{equation}
where $j^2=-1$, and $M=32$ is the number of frames used for the reconstruction.  In
Eq.\ref{eq2}, the coordinates $x,y$ (with $0<x,y<511$) are integers, which represent the
pixel location within the CCD plane. The pixel size is then the physical CCD pixel size,
i.e. 16 $\mu$m.

\subsection{Holographic reconstruction}
\label{sec:reconstruction}

The problem of the  reconstruction in the context of holographic microsopy has been
discussed in details by \cite{colomb2006automatic} and \cite{colomb2006}. Nevertheless,
the Colomb et al. method refers implicitly to the phase-contrast imaging of
\cite{marquet2005dhm}, and is not well suited to the reconstruction of the 3D image of a
wave-field as done by \cite{grilli2001whole}. Here, to get a 3D image of the wave-field
scattered by the gold particle, we have used a slightly different reconstruction method,
which  is inspired from the reconstruction method used by \cite{mann2005}.

First, we considered  that the measured hologram represents the field
$\emph{E}(x,y,z_0)$ within the CCD conjugate plane $z=z_0$ with respect to the Microscope
Objective (MO), i.e., the plane, whose image is on focus on the CCD detector. Since we
image the sample through a microscope objective MO, we must  compensate the phase
curvature, the phase tilt and the enlargement factor that are related to the presence of
MO as disccused by  \cite{colomb2006}. We have thus:
\begin{equation}\label{EQ_2b}
     \emph{E}(x,y,z=z_0)\;=\; e^{j (K_x x + K_y y)} \; e^{j A (x^2 + y^2) }  \sum_{n=0}^{M}  j^{~n} \emph{I}_{n}
\end{equation}
where $(K_x,K_y)$ and $A$ are the  tilt and lens parameters respectively that must be
determined.

We measured these parameters (or compensated their effect) by an original method that
consists in reconstructing the image of the microscope objective output pupil by the one
Fourier transform reconstruction method of \cite{schnars1994drh}. The lens parameters $A$
is then close to the lens parameter that is used in the pupil reconstruction by the
\cite{schnars1994drh} method. On the other hand the tilt  parameters $(K_x,K_y)$ are
compensated by translating the pupil \cite{schnars1994drh} image in the center of the
calculation grid.

%The magnification factor of the conjugate plane  is measured

The properly compensated measured hologram represents then the field
$\emph{E}(x,y,z=z_0)$ in the conjugate plane $z_0$.  Then, as done by \cite{mann2005} in
holographic microscopy, the field $\emph{E}(x,y,z)$ in the vicinity of the conjugate plane
(i.e. for $z \simeq z_0$) is  calculated by the angular spectrum method, which involves
two Fourier transforms (see for example  \cite{Leclerc2000}, \cite{leclerc2001} or
\cite{yu2005}). This method is chosen here since it keeps the pixel size $\delta z$
constant whatever the reconstruction distance $z$ is.

The pixel size $\delta x=\delta y$, which must be calibrated to make a quantitative
analysis of the holographic data,  is measured by imaging with the same setup geometry a
calibrated USAF target located in the CCD conjugate plane. We get $\delta x=\delta y=177$
nm. The reconstruction is then done for 512 different reconstruction distances
\begin{equation}\label{EQ_2c}
  z=z_0+ (m_z-256) \delta z
\end{equation}
where $\delta z=177$ nm and  $m_z=0...511 $. By  this way, we get 3D volume  images with
$512 \times 512\times 512$ voxels, with the same pixel size ($\delta x= \delta y=\delta
z=177$ nm) in the 3 directions $x,y$ and $z$.

\section{Results and Discussion}
\label{sec:}
We have studied different
samples of gold marked cell, unmarked cells and of free 40 nm gold particles.

\subsection{Samples of marked cells}

\begin{figure}[h]
\begin{center}
   \begin{tabular}{c}
   \includegraphics[height=5cm]{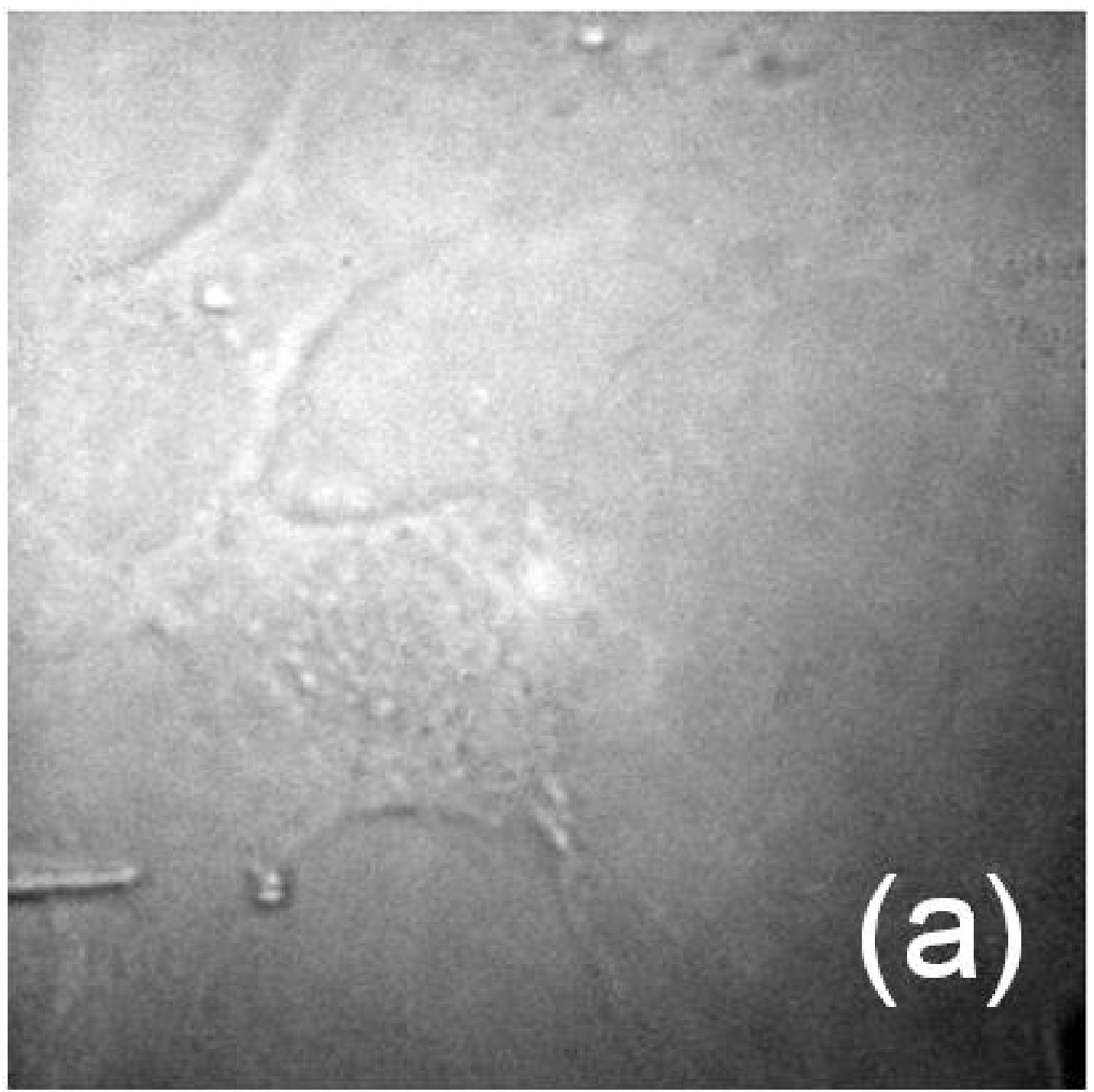}
   \includegraphics[height=5cm]{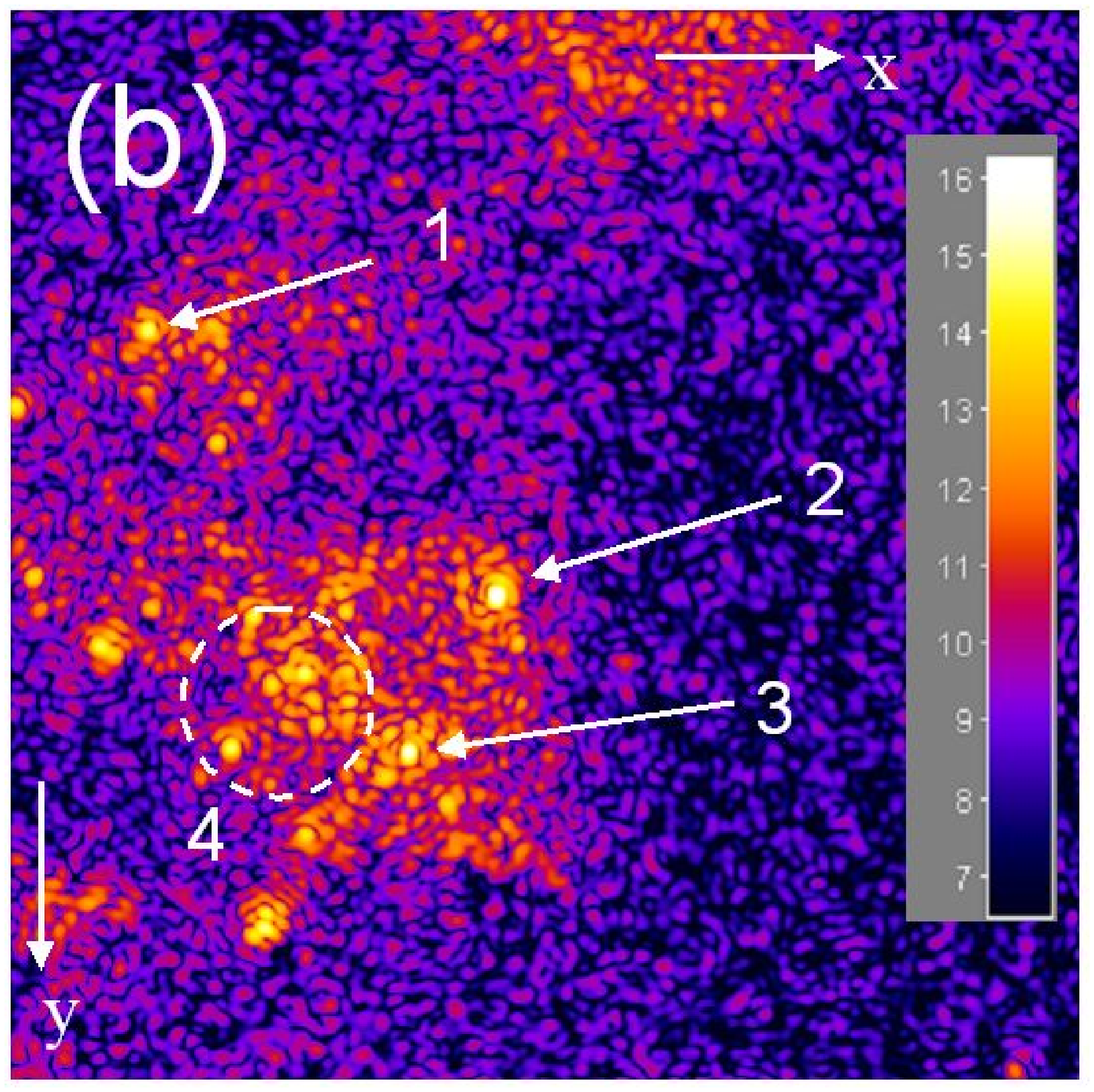}
   \includegraphics[height=5cm]{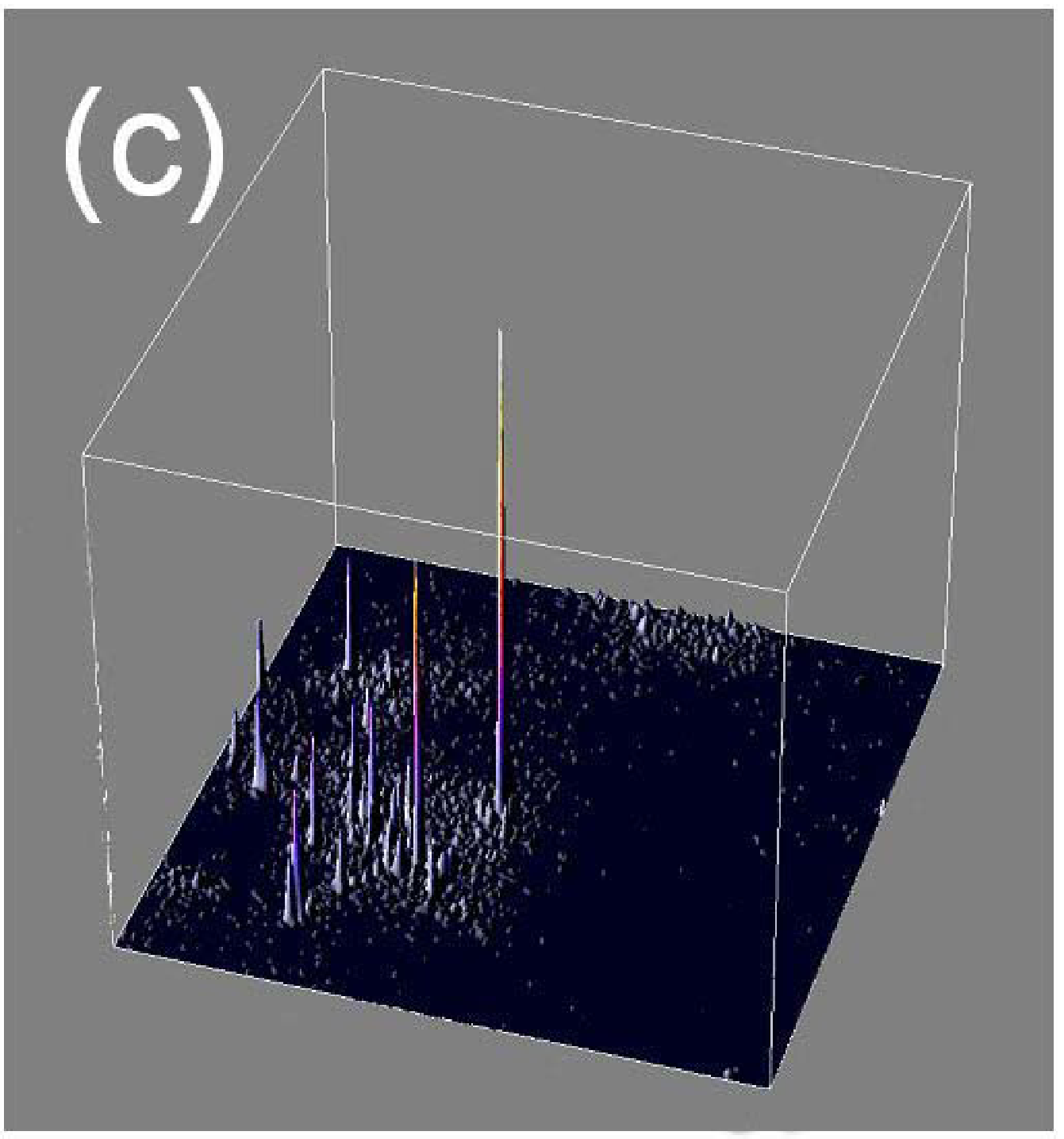}
   \end{tabular}
  \end{center}
\caption{Images of the first sample  with two fibroblast cells marked with 40 nm gold particles.
(a) Direct white light illumination image. (b) Holographic intensity reconstructed image near the
$z = z0$ conjugate plane (z = 255). The display is made in linear grey scale for the intensity
$I = |E|^2$. Black correspond to $\ln I = 5.9$, white to $\ln I = 16.3$ Digital Counts (DC).
(c) 3D linear surface plot of the same data. }
\label{fig_3}       % Give a unique label
\end{figure}

Figures \ref{fig_3} show the images of a first sample, with two cells and several particles.
Figure  \ref{fig_3}(a) shows a direct white light illumination image of the sample. The two
cells can be seen, but the contrast is low. Figure  \ref{fig_3} (b) shows the intensity
holographic image of the same region of the sample reconstructed in  the $z = z_0$ or
$z=256$ conjugate plane (here and in the following we will express the $x$, $y$ and $z$
coordinates by the corresponding pixel index $m_x$, $m_y$ and $m_z$). The display is made
in logarithmic color scale.

Because of the variation of the refractive index within the cells, the illumination light
is diffracted yielding a speckle pattern that is superimposed with the particle signal.
This speckle is visible on Fig. 3 (b), and, from the envelope of the speckled zone, one
can guess the shape of the cells. We interpret the brightest points 1, 2 and 3 of Fig.
\ref{fig_3} (b), which correspond to maximum intensities $I = 11.5 \times 10^6$, $4.6
\times 10^6$, and $8.4\times 10^6$  and Digital Count (DC) respectively, as being
particles signal. Many other bright points are also visible, but it is not simple to
determine, which points are particles, and which are speckle hot spots. This is
especially true within circle 4, where many bright points, close together, are visible.

To better visualize the 40 nm gold particles, we have displayed, on Fig. \ref{fig_3} (c),
by using the Interactive 3D Surface Plot plug-in of Image J (see
\cite{abramoff2004image}), a 3D linear surface plot  of the region of the sample that is
displayed on Fig.\ref{fig_3} (b). As seen, the particles that correspond to sharp peaks
can be easily visualized, but some ambiguity remains around the meaning of the
lower peaks, which could be attributed either to particles, noises, or scattering by biological features of the cell.

\begin{figure}[h]
\begin{center}
   \begin{tabular}{c}
   \includegraphics[height=5cm]{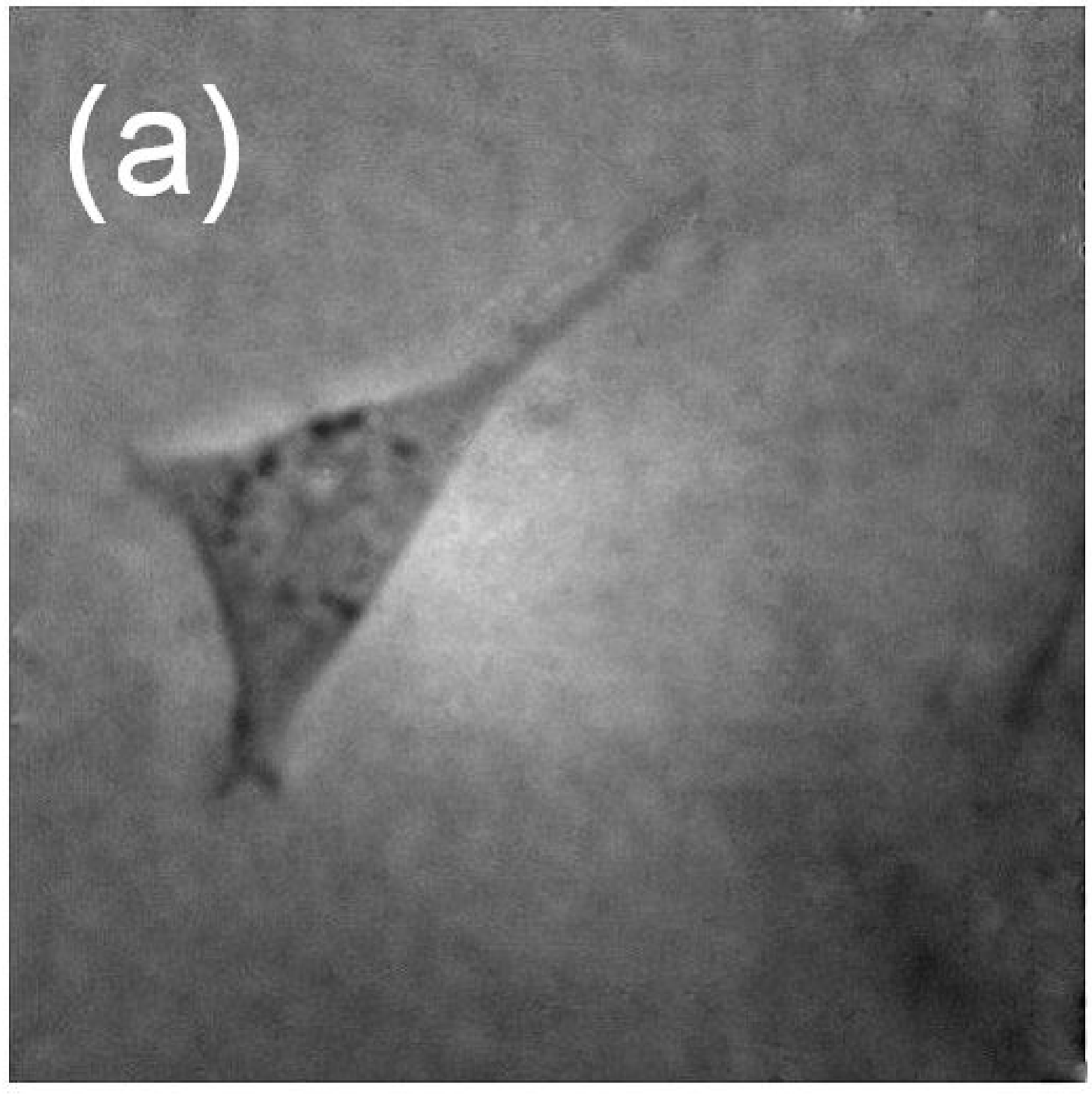}
   \includegraphics[height=5cm]{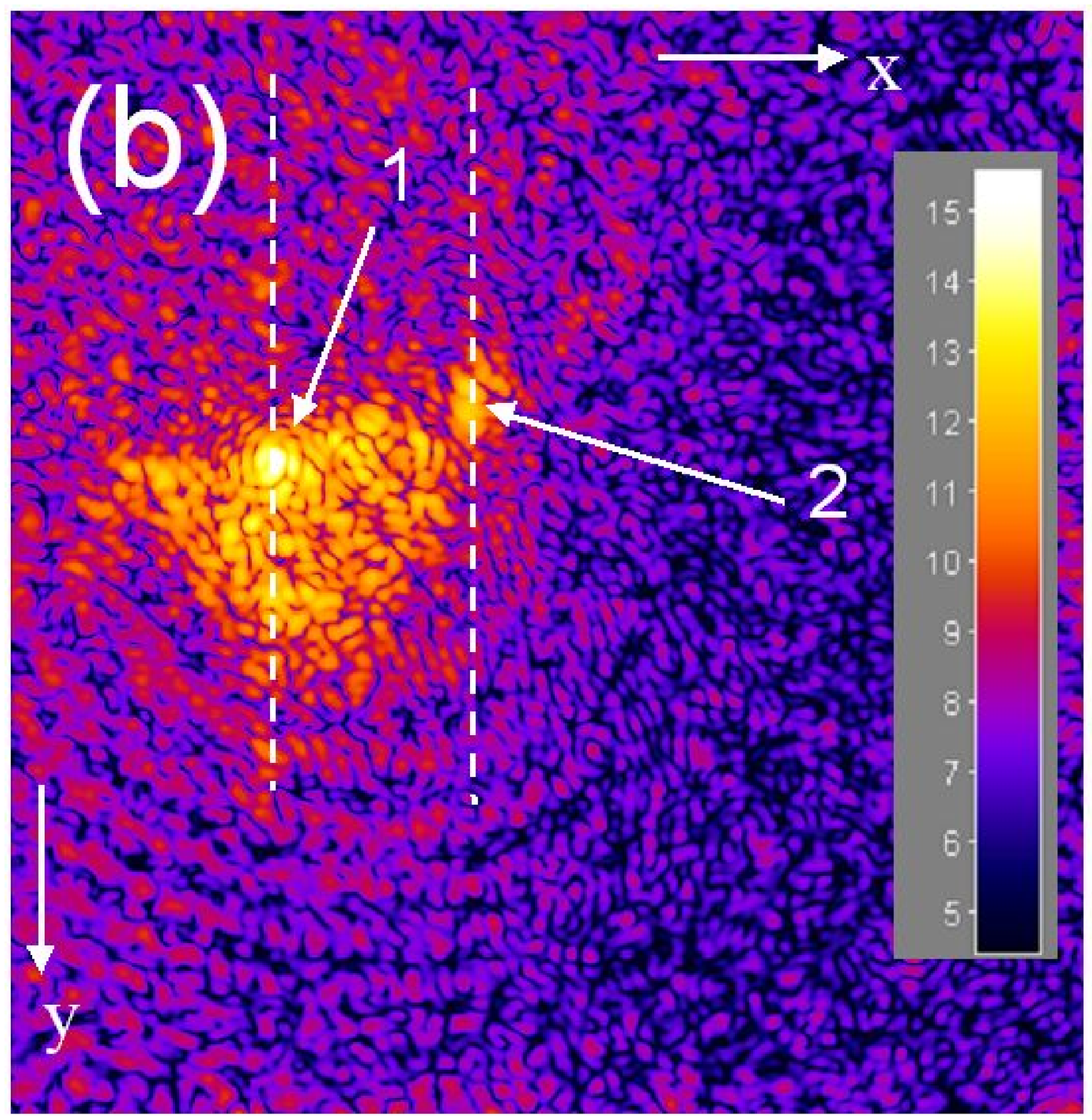}
   \includegraphics[height=5cm]{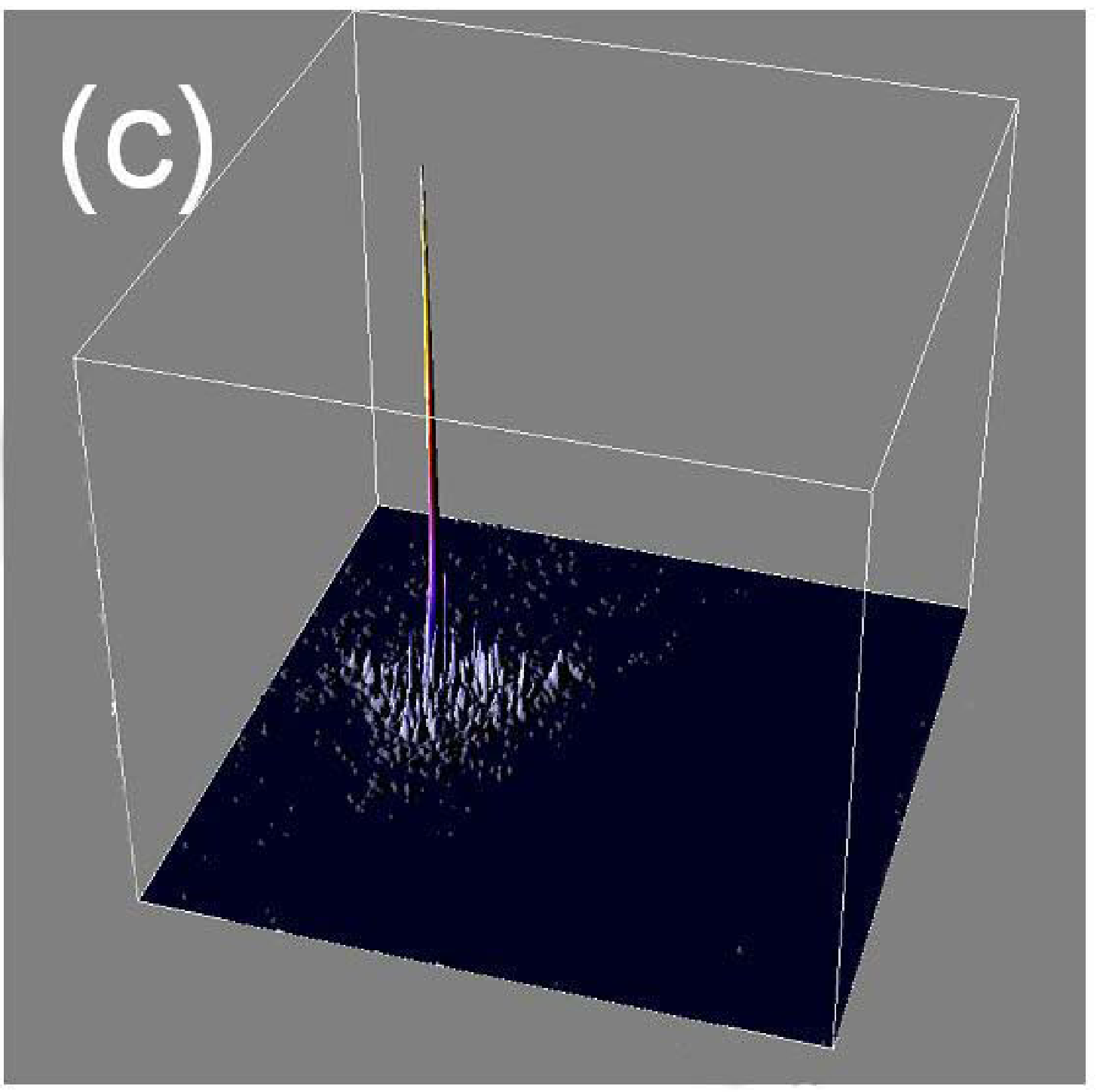}
    \end{tabular}
  \end{center}
\caption{Images of the second sample  with one fibroblast cells marked with one 40 nm gold particle.
(a) Direct white light illumination image. (b) Holographic intensity reconstructed image for the
$z = 325$. The display is made in logarithmic color scale  for the intensity
$I = |E|^2$. Black correspond to $\ln I = 4.48$, white to $\ln I = 15.56$ Digital Counts (DC).
(c) 3D linear surface plot. }
\label{fig_4}       % Give a unique label
\end{figure}

The  images of Fig.\ref{fig_4} are obtained for a second sample with a single cell, and, as
we will see, a single gold particle. Figure \ref{fig_4} (a) shows a white light image of
the sample. One can see the cell, whose shape is triangular. Figures  \ref{fig_4} (b)
shows the intensity holographic image reconstructed for $z = 325$ (i.e. 6.93 $\mu$m above
the $z = z_0$ conjugate plane), with logarithmic color scale display. Here again, the
speckle related to the light diffracted by the cell is visible on Fig.\ref{fig_4} (b),
and one can guess the triangular shape of the cell. The brightest point (arrow 1 on Fig.
\ref{fig_4} (b)) is interpreted as a particle. Since the illumination intensity and focusing
area is not well controlled, the particle maximum intensity $I = 2.9 \times 10^6$ DC
obtained here is noticeably lower than for the first sample. Nevertheless, the signal
obtained for the bright point marked by arrow 2 is more than 10 times lower (i.e. $I =
1.9\times 10^5$), so we can interpret it as a speckle hot spot. This result is confirmed
by Fig. \ref{fig_4} (c) that shows a 3D linear surface plot of the sample.

\begin{figure}[]
\begin{center}
   \begin{tabular}{c}
   \includegraphics[height=6cm]{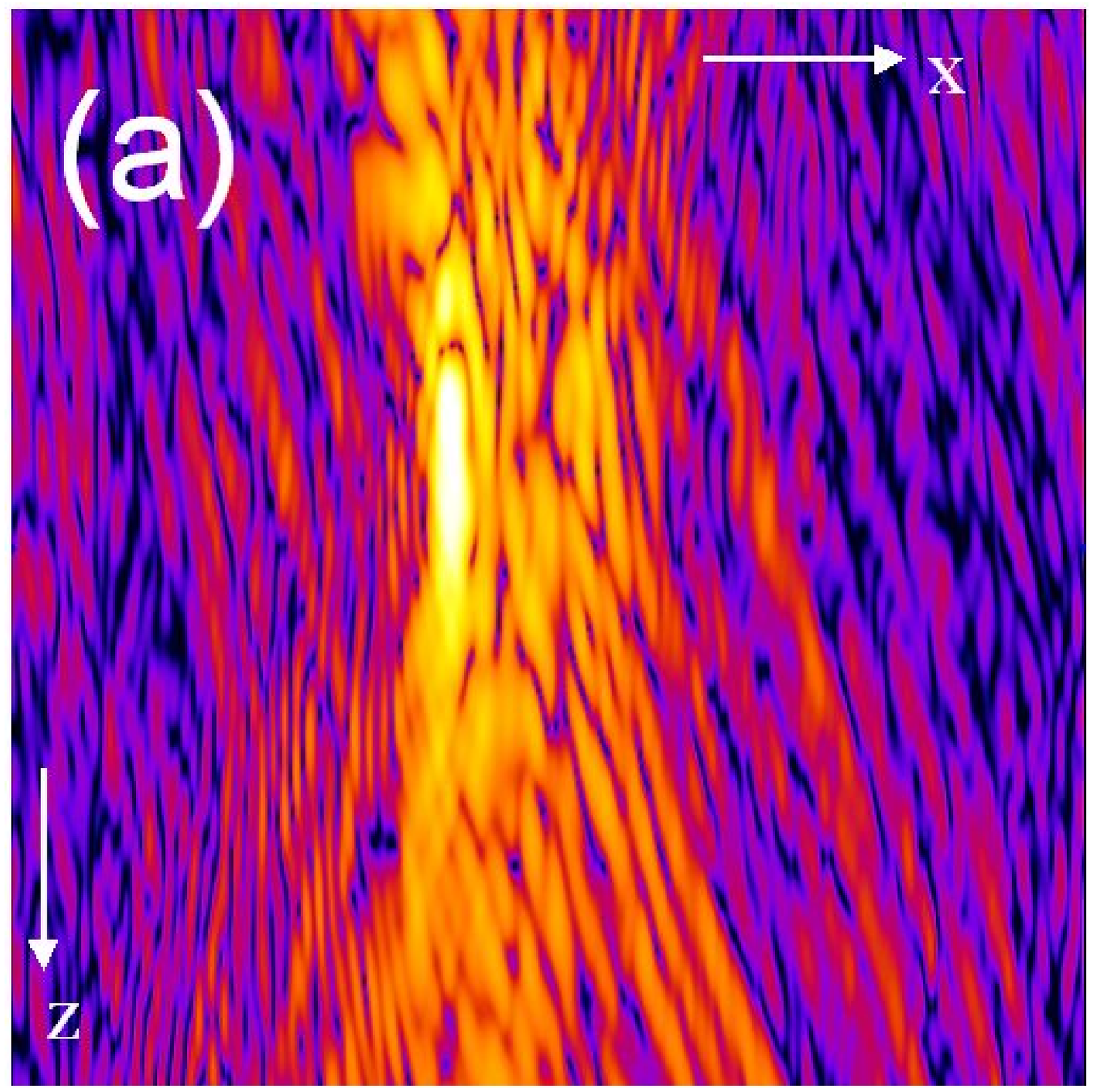}
   \includegraphics[height=6cm]{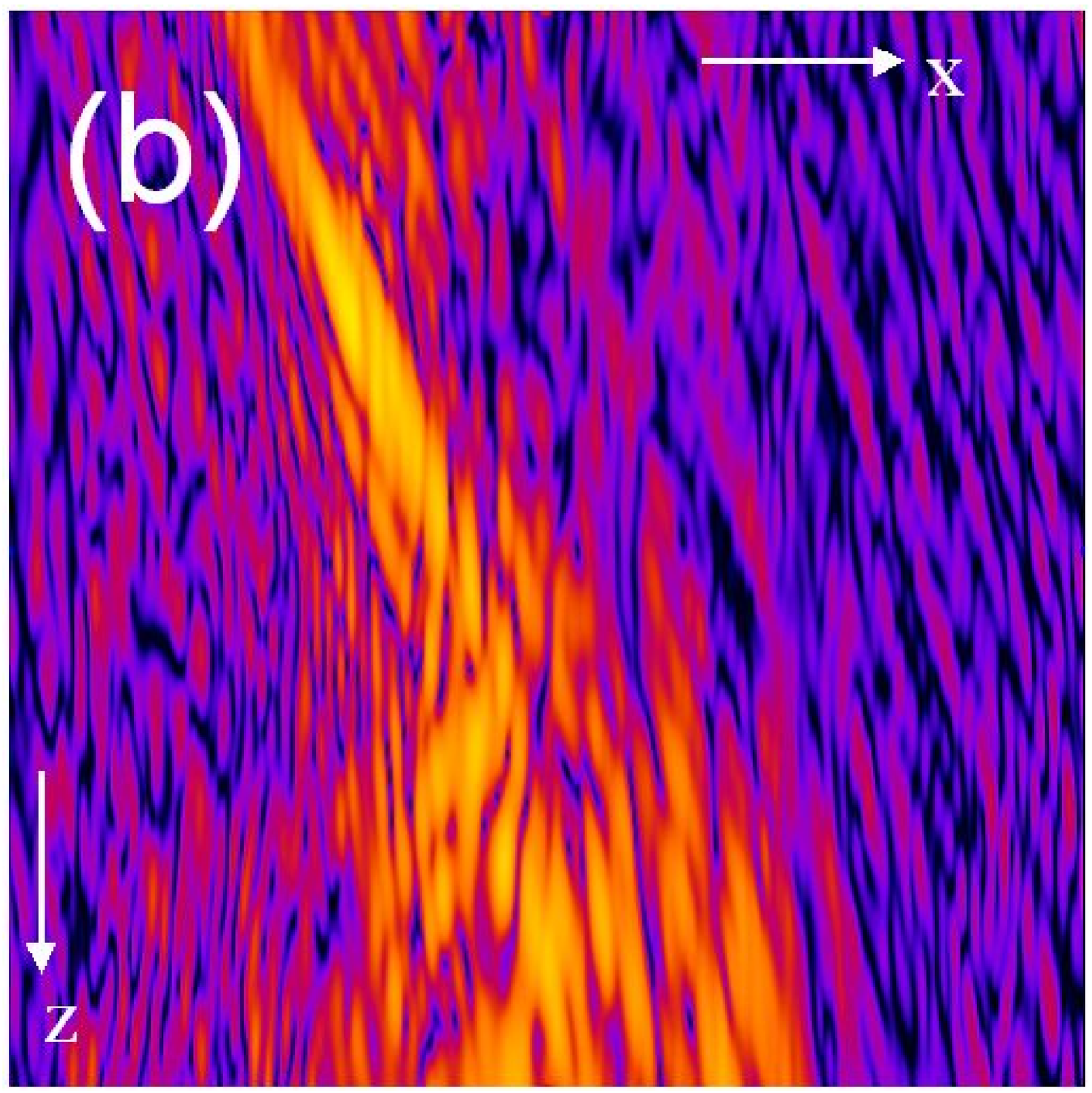}
    \end{tabular}
  \end{center}
\caption{
Images of the second sample obtained by performing cuts of the 3D data  parrallel to the $y$ and $z$ axis. Cuts are made
in planes $x=128$ (a) and $x=221$ (2), which corresponds to the white dashed lines 1 and
2 of Fig.\ref{fig_4}(b) respectively. The display is made  in the same logarithmic color scale
than  Fig.\ref{fig_4}(b) by Volume Wiever. }
\label{fig_5}       % Give a unique label
\end{figure}

To go further, and to better characterize the particle's signal with respect to hot spots,
we have analyzed the 3D images of the wave-field obtained by performing the holographic
reconstruction for the 512 different reconstruction distances of Eq.\ref{EQ_2c}. The 3D
data were displayed using the Volume Viewer plug-in of Image J, which is able to display cuts
of 3D data in arbitrary planes.
By using this plug-in, we have performed cuts parallel to the $yz$ plane of incidence of
the sample illumination beam (see Fig.\ref{fig_2setup} (b) ).

Figure \ref{fig_5} (a) shows the image of a cut made in the plane  $x=128$, which
intersects the particle signal 1 of Fig.\ref{fig_4} (b). The particle is seen as a bright
spot on the $yz$ cut image. We can notice here that the particle signal is located at $z
\simeq 325$ coordinate, which corresponds to the reconstruction plane of Fig.\ref{fig_4}
(b). This is expected, since we have chosen to display on Fig.\ref{fig_4}  (b) the plane
where the maximum intensity is reach in 3D, and since this maximum correspond to the
particle we consider here.

The image of Fig.\ref{fig_5}  (b) corresponds to $x=221$ i.e. to a cut that intersects the
bright spot 2 of Fig.\ref{fig_4}  (b) we interpret as a speckle hot spot. Contrarily to
the particle, the hot spot signal extension along the $x$ axis (vertical axis on the
Fig.\ref{fig_5} images) is quite large. Moreover, the hot spot image is angularly tilted
in the $yz$ plane.

This angular tilt can be simply interpreted by describing how the light propagation is governed in the biological cells. This propagation is dependant on the phase function inside the illuminated cell. Because biological tissues are inhomogeneous, the form of their phase function is not well defined and is thus characterized by the anisotropy coefficient g, which is the average cosine of the phase function. This parameter g describes the asymmetry of the single scattering pattern ; it is thus null when the scattering is isotropic, equals 1 for forward scattering and equals -1 in the case of backward scattering.
In our experiments, the illuminated cells are maintained in DMEM medium that consists mainly of water. Since the refractive index of cells is close to that of water, the cell anisotropy factor g is close to one ($g \simeq 0.9$ in biological tissues as mentioned by \cite{cheong1990review}). As a consequence, the light scattered by the cells mainly follows the forward scattering regime and the observed light scattering pattern appears to be tilted by approximately 45$^\circ$ since the incident illumination laser beam is initially tilted by 45$^\circ$ in conformity with the TIR illumination geometry (see Fig.\ref{fig_2setup} (b)).

The exact shape of the hot spot's wave-field can be calculated, but it is
quite complicated. It involves the calculation of the angular distribution of the
scattered light, which depends on the cell anisotropy factor $g$. One must then calculate
the refraction of the scattered light on the medium-air interface, and take into consideration
the collection of light by the microscope objective. The quantitative analysis of
the wave-field's shape, which yields the angular tilt, is thus out of the scope in the present
paper, and one can simply say that the hot spot signal keeps some memory of the
illumination direction, and is thus angularly tilted in the $y z$ plane.

One can notice that a similar angular tilt effect has been observed recently on the
photothermal signal of 50 nm and 10 nm gold particles by \cite{absil2010photothermal}.

\subsection{Control experiments performed on free particles or on unmarked-cells}

In order to confirm our interpretation of the angular tilt seen on Fig.\ref{fig_5} (b),
we have performed some control experiments by imaging an unmarked cell sample and another sample of free gold particles.

\subsubsection{Unmarked cell sample }

\begin{figure}[]
\begin{center}
   \begin{tabular}{c}
   \includegraphics[height=5cm]{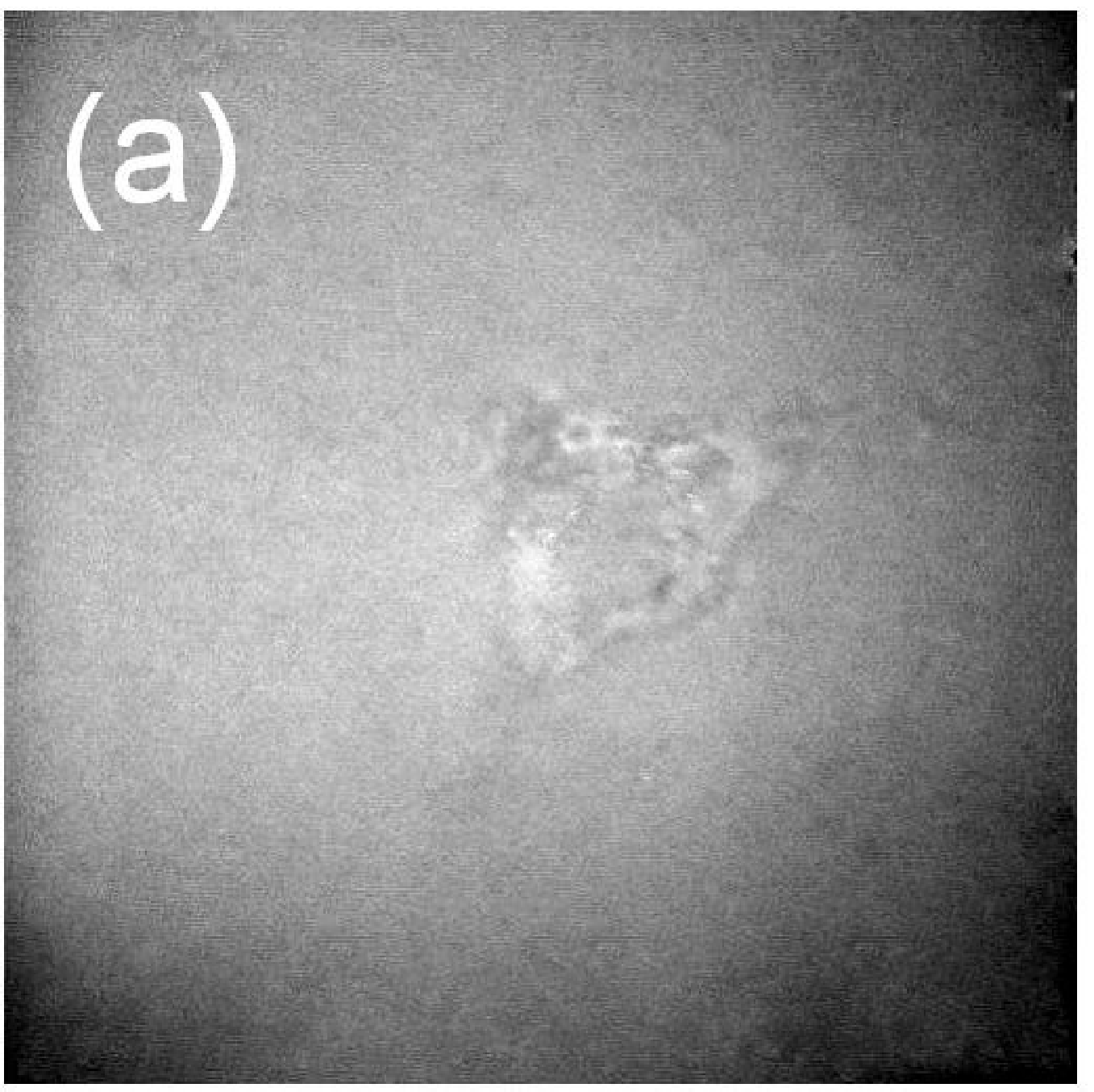}
   \includegraphics[height=5cm]{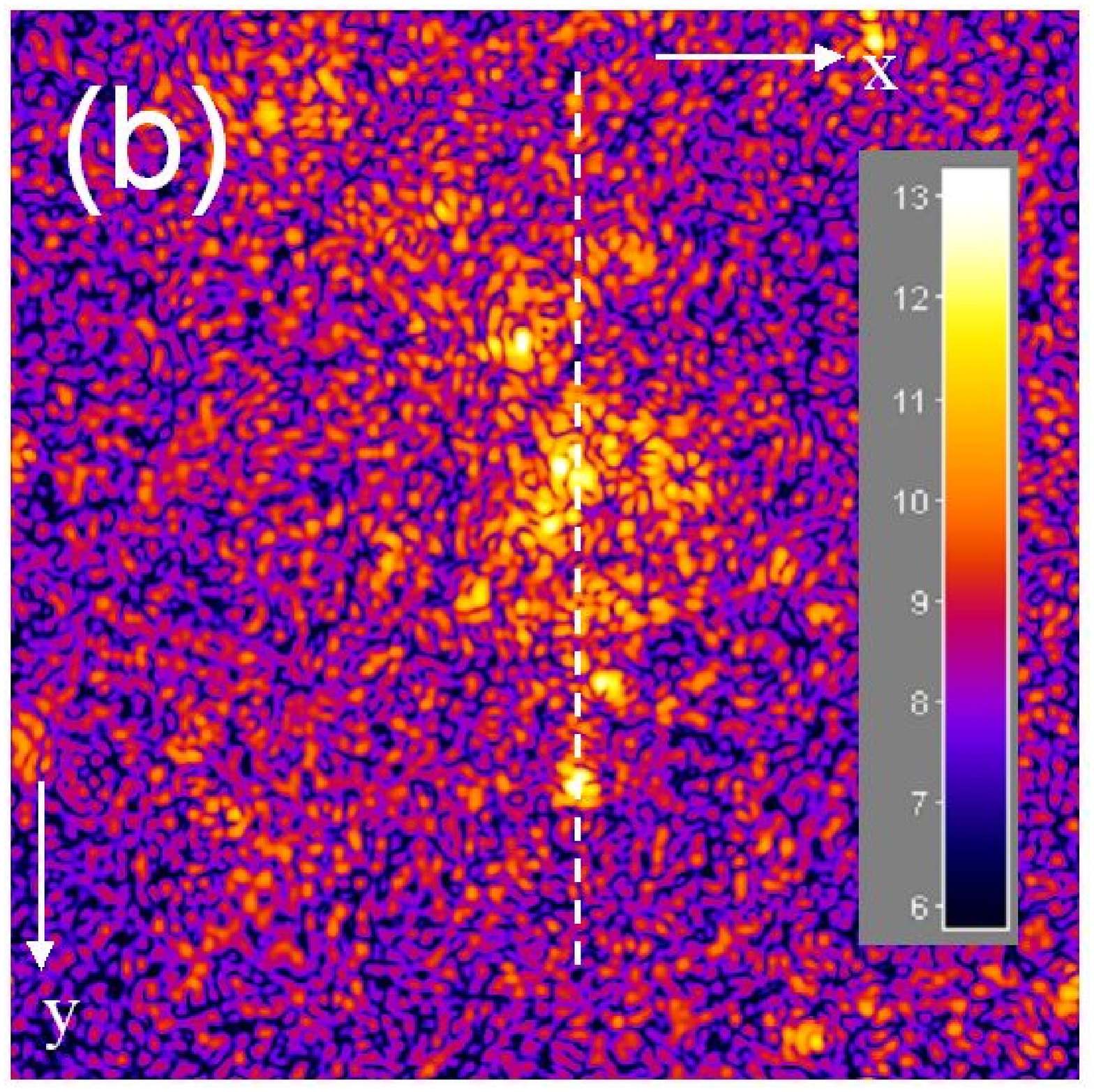}
  \includegraphics[height=5cm]{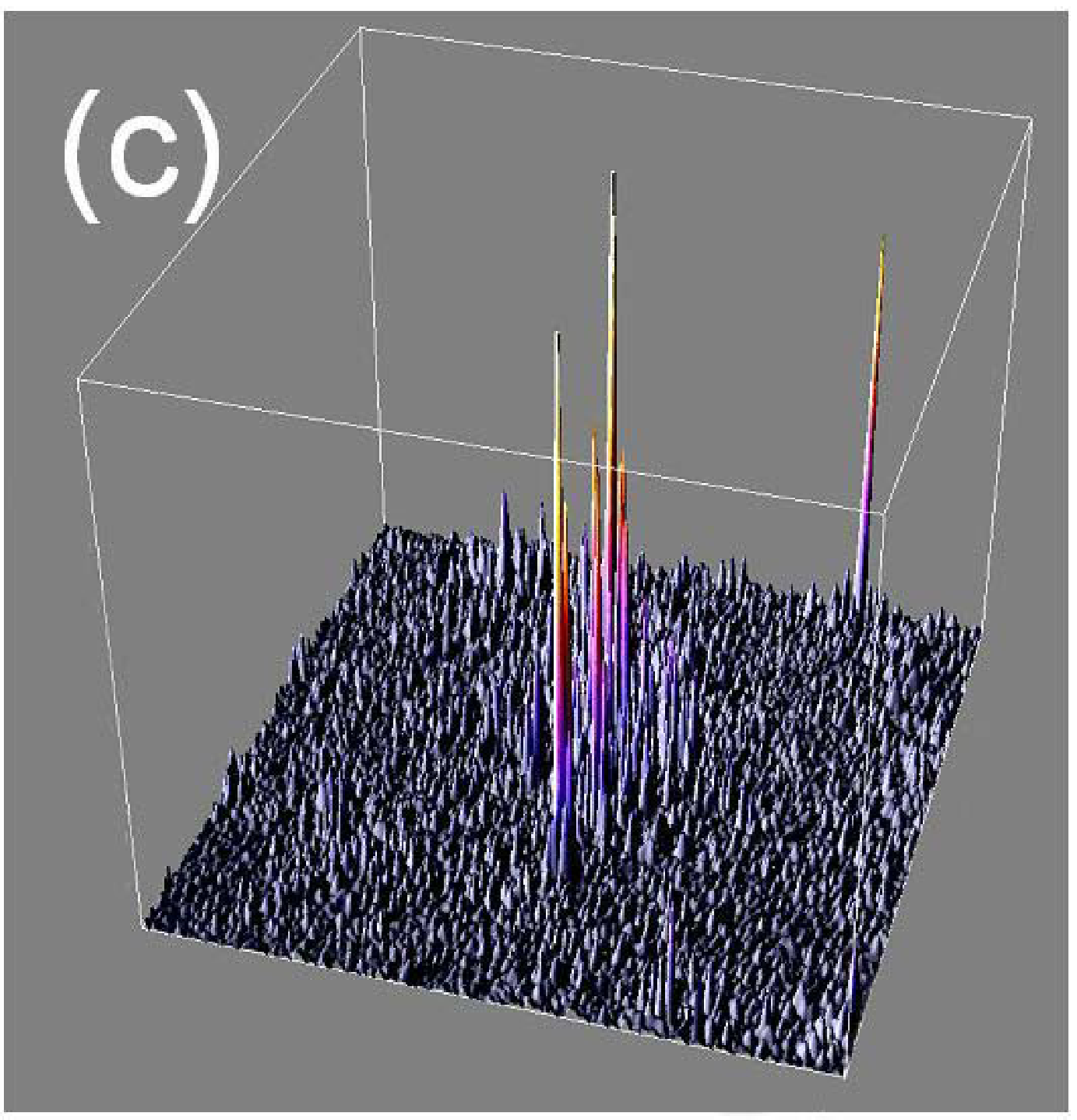}\\
   \includegraphics[height=6cm]{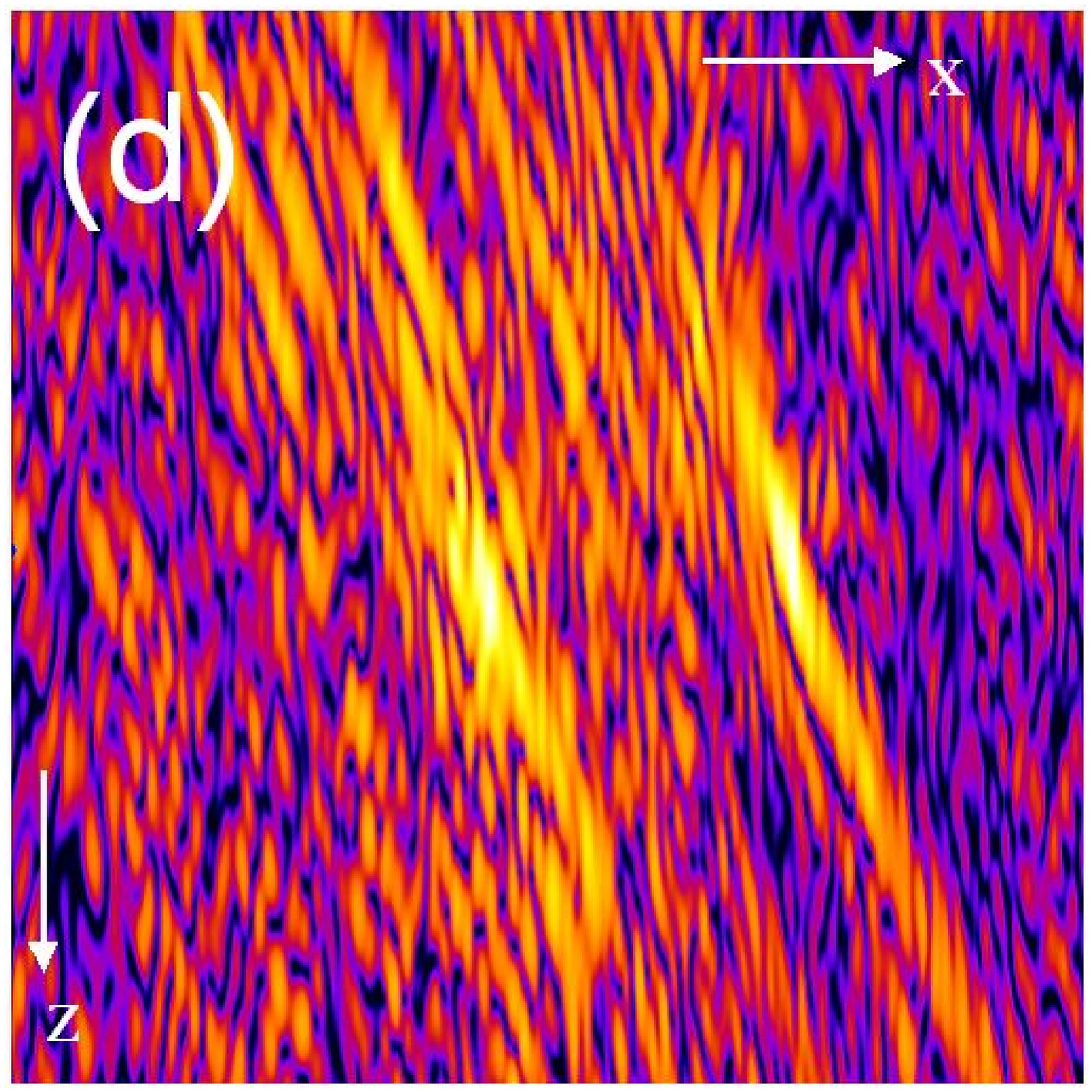} \\
    \end{tabular}
  \end{center}
\caption{
Images of an unmarked cell sample. (a) Direct white light illumination image. (b) Holographic intensity reconstructed image for the
$z = 252$. The display is made in logarithmic color scale  for the intensity
$I = |E|^2$. Black correspond to $\ln I = 5.80$, white to $\ln I = 13.28$ Digital Counts (DC).
(c) 3D linear surface plot. (d) Image obtained par performing cut of the 3D data
in the plane $x=271$ that corresponds to the white dashed line seen in (b). The display is made by Volume Viewer with the same logarithmic
color scale as in (b).}
\label{fig_6}       % Give a unique label
\end{figure}

Fig.\ref{fig_6} shows the results  obtained  with a cell sample without particle. The
direct white light image is shown in Fig. \ref{fig_6}(a). Figure \ref{fig_6}(b) shows the
holographic intensity reconstructed image that shows the light scattered  by the cell
refractive index inhomogeneities. This scattered light, which has a speckle structure,
exhibits several hot spots that correspond to bright points on Fig.\ref{fig_6}(b), and we
have chosen to image on Fig.\ref{fig_6}(b) the  reconstruction plane $z=252$ that
corresponds to the maximum hot spot intensity ($5.85\times 10^5$ DC).

The image we get here without particles is visually quite similar to Fig.\ref{fig_3}(b) which
is obtained with several particles. The 3D linear surface plot shown in
Fig.\ref{fig_3}(c) does not help since it is visually similar to Fig.\ref{fig_3}(c).
Nevertheless,   the maximum of the bead signal ($11.6\times 10^6$ DC) in
Fig.\ref{fig_3}(b) and (c) is much higher than the maximum of the hot spot signal
($5.85\times 10^5$ DC) in Fig.\ref{fig_6}(b) and (c). This makes the background noise
visually higher in Fig.\ref{fig_6}(c).

Figure \ref{fig_6}(d) shows the $yz$ image obtained by performing a cut in the plane
$x=271$ that intersects the two brightest speckle hot spots of the sample. Now the  $yz$
image obtained in Fig. \ref{fig_6}(d) without particle  is qualitatively different than
with a particle in Fig. \ref{fig_5}(a). The signal is angularly tilted like in Fig.
\ref{fig_5}(b), and its extension in the $z$ direction (vertical axis) is larger.

\subsubsection{Free particles sample}

\begin{figure}[]
\begin{center}
   \begin{tabular}{c}
   \includegraphics[height=5cm]{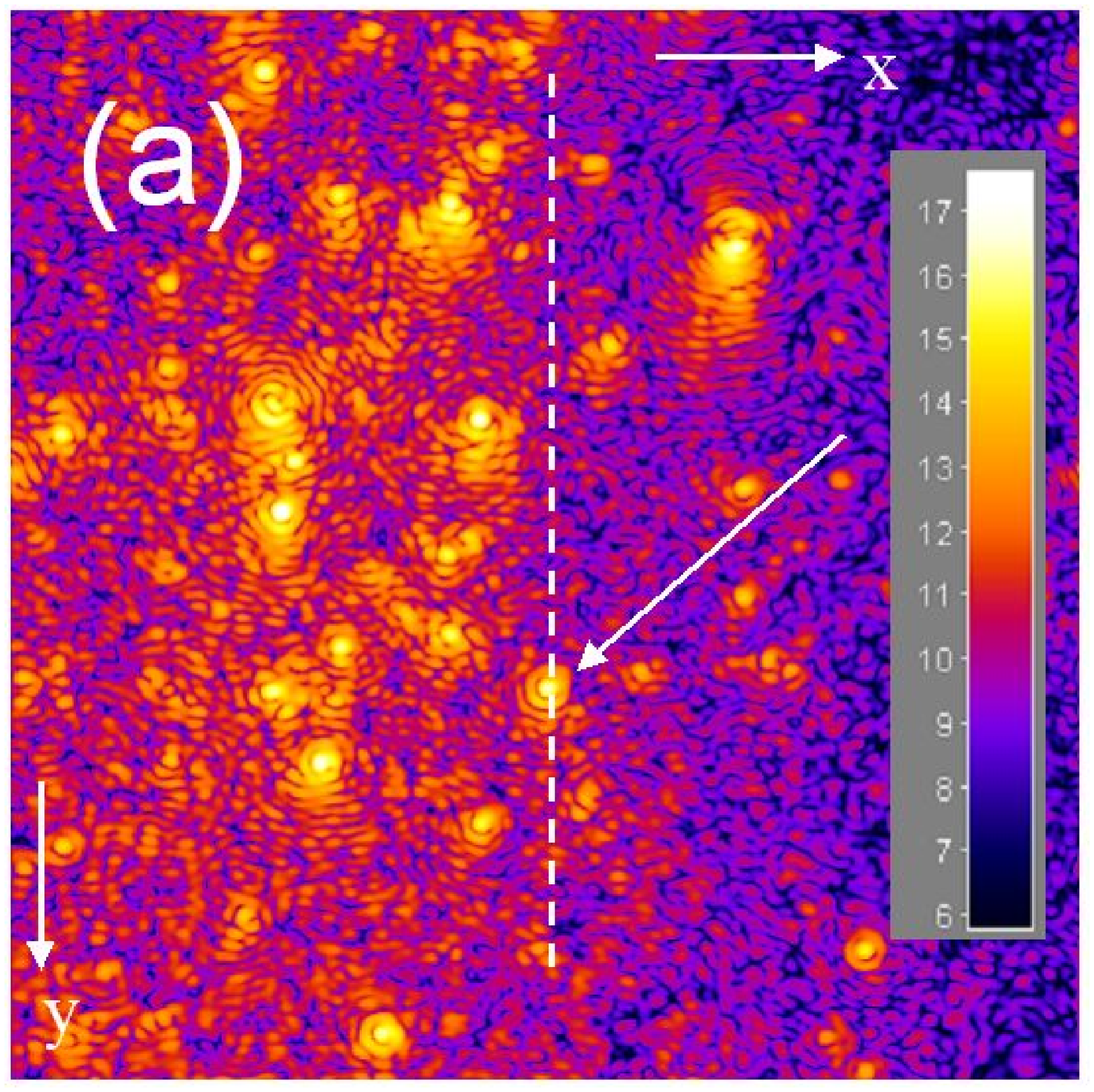}
   \includegraphics[height=5cm]{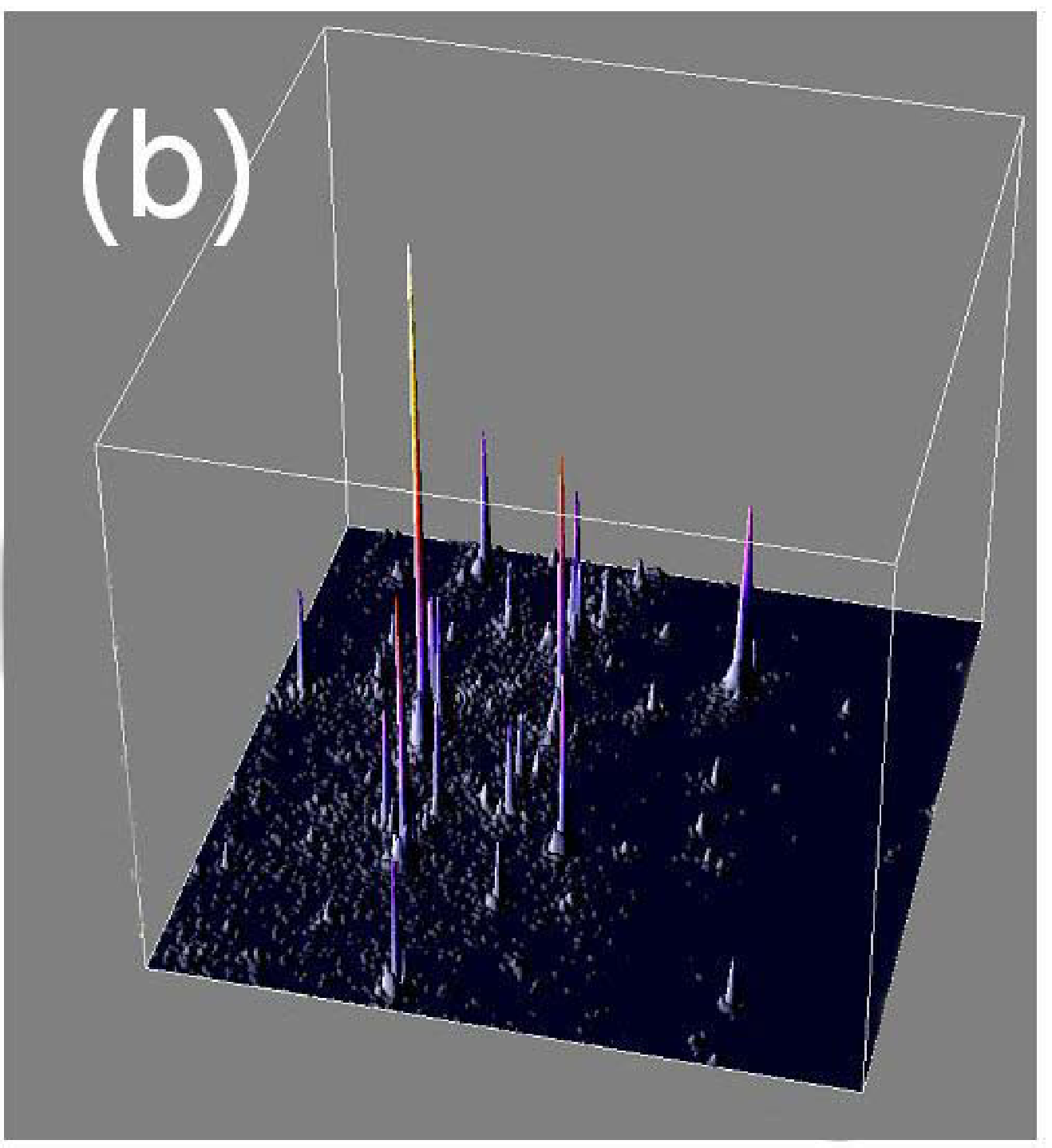}\\
  \includegraphics[height=6cm]{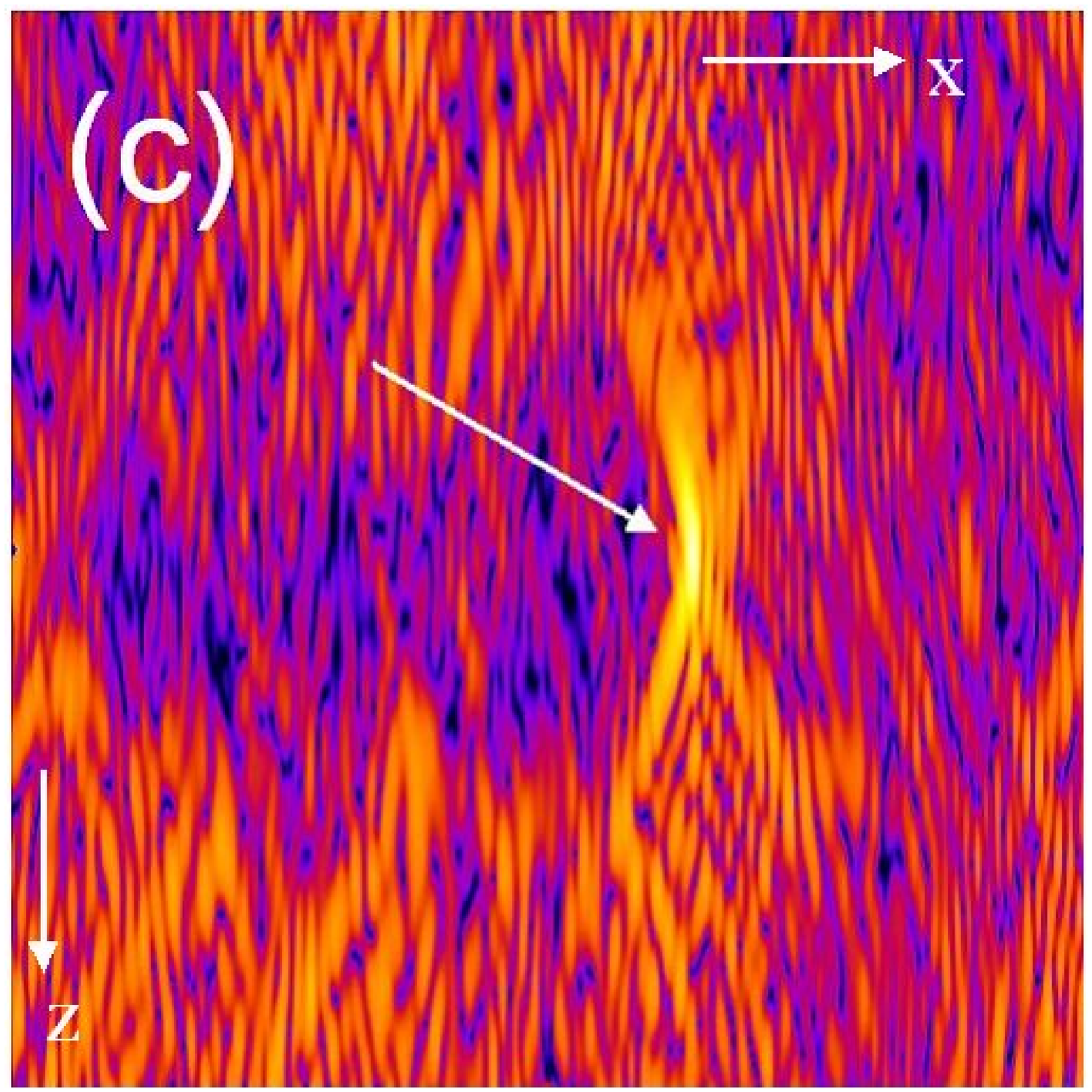}\\
     \end{tabular}
  \end{center}
\caption{
Images of free gold particles in a water and agarose suspension.  (a) Holographic intensity reconstructed image for the
$z = 251$. The display is made in logarithmic color scale  for the intensity
$I = |E|^2$. Black correspond to $\ln I = 5.85$, white to $\ln I = 17.61$ Digital Counts (DC).
(b) 3D linear surface plot. (c) Image obtained par performing cut of the 3D data
in the plane $x=258$ that corresponds to the white dashed line seen in (a).
The display is made by Volume Viewer with the same logarithmic
color scale than for (a).}
\label{fig_7}   % Give a unique label
\end{figure}

The images of Fig.\ref{fig_7} are obtained for free particles (uncoupled to cells) in brownian
motion in a water and agarose suspension. Here, we used agarose to slow down the motion
of the particles enough to make the displacement of the particles negligible during the
time of acquisition of the sequence of $M=32$ frames.

Figure \ref{fig_6}(a) is the holographic intensity reconstructed image. The brightest
points  correspond to particles that are located within the reconstruction plane, while
the less bright points are interpreted as out of focus particles. Reconstruction is made
here in the plane $z=251$ which corresponds to the maximum intensity ($4.4\times 10^7$ DC) for
the brightest spot (marked by a white arrow on Fig.\ref{fig_6}(a)). Many particles can be
seen on Fig. \ref{fig_6}(a) and on Fig. \ref{fig_6}(b) that displays the holographic
data with 3D linear surface plot.

Figure \ref{fig_6}(c) shows the $yz$ image in the plane $x=258$ that intersects the
brightest particle and corresponds to the white dashed line on Fig.\ref{fig_6}(a). As
expected,  the image of the particle (white arrow) has a smaller extension along the $z$
direction (vertical direction) and does not exhibit the angular tilt we get with speckle
hot spot in Fig.\ref{fig_5}(b) and Fig.\ref{fig_6}(d). This result confirms our
interpretation of the angular tilt observed with the speckle hot spots.

\section{Conclusion}
\label{conc}

To summarize, we have shown that heterodyne holographic microscopy, in the off-axis geometry, is well
adapted to the detection of weakly scattering objects. The sensitivity,
signal to noise ratio and selectivity of the technique allow the
localization of light-scattering gold nanoparticles of a few tens of
nanometers, which are good candidates of non-toxic and perennial
markers. Biological environments, however, are difficult to address
since cell features generate strong parasitic speckle. Here, we have reported
the detection of 40 nm particles attached to the surface of live 3T3
mouse fibroblasts. A comparison of these signals with either
non-labelled cells or simple gold particles in solution allowed us to
unambiguously discriminate particles. We show that, in addition to a
stronger scattering signal, gold particles induce a relatively isotropic
scattering, whereas biological features are characterized by mostly
forward scattering.
This dissimilarity in the scattering patterns, explained by the inconsistency of the refractive
indexes and anisotropy parameters g, is easily characterized by
digital holography, making it an excellent tool for the 3D detection
of gold labels in biological environments.

\begin{acknowledgements}
Authors wish to acknowledge the French Agence Naionale de la Recherche (ANR) and the "Centre
de Comp\'{e}tence NanoSciences Ile de France" (C'nano IdF) for their financial support.
\end{acknowledgements}

%\bibliography{bibtheseFadwa}   % name your BibTeX data base

\begin{thebibliography}{22}
\providecommand{\natexlab}[1]{#1}
\providecommand{\url}[1]{{#1}}
\providecommand{\urlprefix}{URL }
\expandafter\ifx\csname urlstyle\endcsname\relax
  \providecommand{\doi}[1]{DOI~\discretionary{}{}{}#1}\else
  \providecommand{\doi}{DOI~\discretionary{}{}{}\begingroup
  \urlstyle{rm}\Url}\fi
\providecommand{\eprint}[2][]{\url{#2}}

\bibitem[{Abramoff et~al(2004)Abramoff, Magalhaes, and Ram}]{abramoff2004image}
Abramoff M, Magalhaes P, Ram S (2004) {Image processing with ImageJ}.
  Biophotonics International 11(7):36--43

\bibitem[{Absil et~al(2010)Absil, Tessier, Gross, Atlan, Warnasooriya, Suck,
  Coppey-Moisan, and Fournier}]{absil2010photothermal}
Absil E, Tessier G, Gross M, Atlan M, Warnasooriya N, Suck S, Coppey-Moisan M,
  Fournier D (2010) {Photothermal heterodyne holography of gold nanoparticles}.
  Opt Express 18:780--786

\bibitem[{Atlan et~al(2007)Atlan, Gross, and Absil}]{atlan2007aps}
Atlan M, Gross M, Absil E (2007) {Accurate phase-shifting digital
  interferometry}. Opt Lett 32:1456--1458

\bibitem[{Atlan et~al(2008)Atlan, Gross, Desbiolles, Absil, Tessier, and
  Coppey-Moisan}]{atlan2008heterodyne}
Atlan M, Gross M, Desbiolles P, Absil {\'E}, Tessier G, Coppey-Moisan M (2008)
  {Heterodyne holographic microscopy of gold particles}. Optics letters
  33(5):500--502

\bibitem[{Boyer et~al(2003)Boyer, Tamarat, Maali, Lounis, and
  Orrit}]{boyer2003}
Boyer D, Tamarat P, Maali A, Lounis B, Orrit M (2003) {Photothermal imaging of
  nanometer-sized metal particles among scatterers}. Science 297:1160--1163

\bibitem[{Cheong et~al(1990)Cheong, Prahl, and Welch}]{cheong1990review}
Cheong W, Prahl S, Welch A (1990) {A review of the optical properties of
  biological tissues}. IEEE journal of quantum electronics 26(12):2166--2185

\bibitem[{Cognet et~al(2002)Cognet, C.~Tardin, Boyer, Choquet, Tamarat, and
  Lounis}]{cognet2002}
Cognet L, C~Tardin C, Boyer D, Choquet D, Tamarat P, Lounis B (2002) {Single
  metallic nanoparticles imaging for protein detection in cells}. Proc Natl
  Acad Sci 100:11,350--11,355

\bibitem[{Colomb et~al(2006{\natexlab{a}})Colomb, Cuche, Charri{\`e}re,
  K{\"u}hn, Aspert, Montfort, Marquet, and Depeursinge}]{colomb2006automatic}
Colomb T, Cuche E, Charri{\`e}re F, K{\"u}hn J, Aspert N, Montfort F, Marquet
  P, Depeursinge C (2006{\natexlab{a}}) {Automatic procedure for aberration
  compensation in digital holographic microscopy and applications to specimen
  shape compensation}. Applied optics 45(5):851--863

\bibitem[{Colomb et~al(2006{\natexlab{b}})Colomb, Montfort, Kuehn, Aspert,
  Cuche, Marian, Charri{\'e}re, Bourquin, Marquet, and
  Depeursinge}]{colomb2006}
Colomb T, Montfort F, Kuehn J, Aspert N, Cuche E, Marian A, Charri{\'e}re F,
  Bourquin S, Marquet P, Depeursinge C (2006{\natexlab{b}}) {Numerical
  parametric lens for shifting, magnification, and complete aberration
  compensation in digital holographic microscopy}. J Opt Soc Am A 23:3177--3190

\bibitem[{Goldberg and Burmeister(1986)}]{goldberg1986stages}
Goldberg D, Burmeister D (1986) {Stages in axon formation: observations of
  growth of Aplysia axons in culture using video-enhanced contrast-differential
  interference contrast microscopy}. Journal of Cell Biology 103(5):1921

\bibitem[{Grilli et~al(2001)Grilli, Ferraro, De~Nicola, Finizio, Pierattini,
  and Meucci}]{grilli2001whole}
Grilli S, Ferraro P, De~Nicola S, Finizio A, Pierattini G, Meucci R (2001)
  {Whole optical wavefields reconstruction by digital holography}. Optics
  Express 9(6):294--302

\bibitem[{Jain et~al(2006)Jain, Lee, El-Sayed, and
  El-Sayed}]{jain2006calculated}
Jain P, Lee K, El-Sayed I, El-Sayed M (2006) {Calculated absorption and
  scattering properties of gold nanoparticles of different size, shape, and
  composition: applications in biological imaging and biomedicine}. J Phys Chem
  B 110(14):7238--7248

\bibitem[{Lasne et~al(2006)Lasne, Blab, Berciaud, Heine, Groc, Choquet, Cognet,
  and Lounis}]{lasne2006}
Lasne D, Blab GA, Berciaud S, Heine M, Groc L, Choquet D, Cognet L, Lounis B
  (2006) {Single nanoparticle photothermal tracking (SNaPT) of 5-nm gold beads
  in live cells}. Biophys J 91:4598--4604

\bibitem[{LeClerc et~al(2000)LeClerc, Collot, and Gross}]{Leclerc2000}
LeClerc F, Collot L, Gross M (2000) Numerical heterodyne holography with
  two-dimensional photo-detector arrays. Opt Lett 25:716--718

\bibitem[{LeClerc et~al(2001)LeClerc, Gross, and Collot}]{leclerc2001}
LeClerc F, Gross M, Collot L (2001) {Synthetic-aperture experiment in the
  visible with on-axis digital heterodyne holography}. Opt Lett 26:1550--1552

\bibitem[{Mann et~al(2005)Mann, Yu, Lo, and Kim}]{mann2005}
Mann CJ, Yu L, Lo CM, Kim MK (2005) {High resolution quantitative
  phase-contrast microscopy by digital holography}. Opt Express 13:8693--8698

\bibitem[{Marquet et~al(2005)Marquet, Rappaz, Magistretti, Cuche, Emery,
  Colomb, and Depeursinge}]{marquet2005dhm}
Marquet P, Rappaz B, Magistretti P, Cuche E, Emery Y, Colomb T, Depeursinge C
  (2005) {Digital holographic microscopy: a noninvasive contrast imaging
  technique allowing quantitative visualization of living cells with
  subwavelength axial accuracy}. Optics letters 30(5):468--470

\bibitem[{Schnars and J{\"u}ptner(1994)}]{schnars1994drh}
Schnars U, J{\"u}ptner W (1994) {Direct recording of holograms by a CCD target
  and numerical reconstruction}. Applied Optics 33(2):179--181

\bibitem[{S{\"o}nnichsen et~al(2000)S{\"o}nnichsen, Geier, Hecker, Von~Plessen,
  Feldmann, Ditlbacher, Lamprecht, Krenn, Aussenegg, Chan
  et~al}]{sonnichsen2000spectroscopy}
S{\"o}nnichsen C, Geier S, Hecker N, Von~Plessen G, Feldmann J, Ditlbacher H,
  Lamprecht B, Krenn J, Aussenegg F, Chan V, et~al (2000) {Spectroscopy of
  single metallic nanoparticles using total internal reflection microscopy}.
  Applied Physics Letters 77:2949

\bibitem[{Warnasooriya et~al(2010)Warnasooriya, Joud, Bun, Tessier,
  Coppey-Moisan, Desbiolles, Atlan, Abboud, and
  Gross}]{warnasooriya2010imaging}
Warnasooriya N, Joud F, Bun P, Tessier G, Coppey-Moisan M, Desbiolles P, Atlan
  M, Abboud M, Gross M (2010) {Imaging gold nanoparticles in living cell
  environments using heterodyne digital holographic microscopy}. Opt Express
  18:3264--3273

\bibitem[{West et~al(2006)West, Drezek, and J.}]{west06}
West JL, Drezek RA, J HN (2006) Nanotechnology provides new tools for
  biomedical optics. In: Bronzino JD (ed) Tissue Engineering and Artifical
  Organs, 3rd Edition, CRC Press, pp 25--1--25--9

\bibitem[{Yu and Kim(2005)}]{yu2005}
Yu L, Kim M (2005) {Wavelength-scanning digital interference holography for
  tomographic three-dimensional imaging by use of the angular spectrum method}.
  Opt Lett 30:2092--2094

\end{thebibliography}


%\begin{thebibliography}{}
% etc
%\end{thebibliography}

\end{document}